\documentclass[osajnl,twocolumn,showpacs,superscriptaddress,11pt]{revtex4-1} 
\usepackage{amsmath,amssymb,graphicx}

\usepackage{graphicx}
\usepackage{dcolumn}
\usepackage{bm}
\usepackage{color}
\usepackage{times}
\usepackage{wasysym}
\renewcommand{\vec}[1]{\mathbf{#1}}
\usepackage{graphicx}
\usepackage{times}
\usepackage{multirow}

\newcommand{\vecr}{\ensuremath{\mathbf{r}}}
\newcommand{\epsbar}{\ensuremath{\bar{\epsilon}}}

\renewcommand{\vec}[1]{\mathbf{#1}}

\newcommand{\vE}{\ensuremath{\mathbf{E}}}
\renewcommand{\vr}{\ensuremath{\mathbf{r}}}
\newcommand{\lbda}{\ensuremath{\lambda}}
\newcommand{\w}{\ensuremath{\omega}}
\newcommand{\e}{\ensuremath{\epsilon}}
\newcommand{\lp}{\ensuremath{\left(}}
\newcommand{\rp}{\ensuremath{\right)}}

\renewcommand{\eqref}[1]{(\ref{eq:#1})}

\newcommand{\figref}[1]{Fig.~\ref{fig:#1}}
\newcommand{\Figref}[1]{Figure~\ref{fig:#1}}

\newcommand{\citeasnoun}[1]{Ref.~\onlinecite{#1}}
\newcommand{\citesasnoun}[1]{Refs.~\onlinecite{#1}}

\begin{document}

\title{Topology optimization of multi-track ring resonators and 2D microcavities for nonlinear frequency conversion}

\author{Zin Lin}
\email{Corresponding author: zinlin@g.harvard.edu}
\affiliation{John A. Paulson School of Engineering and Applied Sciences, Harvard University Cambridge, MA 02138}
\author{Marko Lon\v{c}ar}
\affiliation{John A. Paulson School of Engineering and Applied Sciences, Harvard University Cambridge, MA 02138}
\author{Alejandro W. Rodriguez}
\affiliation{Department of Electrical Engineering, Princeton University, Princeton, NJ 08544}

\begin{abstract}
  We exploit recently developed topology-optimization techniques to
  design complex, wavelength-scale resonators for enhancing various
  nonlinear $\chi^{(2)}$ and $\chi^{(3)}$ frequency conversion processes. In
  particular, we demonstrate aperiodic, multi-track ring resonators
  and 2D slab microcavities exhibiting long lifetimes $Q \gtrsim
  10^4$, small modal volumes $V \gtrsim (\lambda/2n)^3$, and among the
  largest nonlinear overlaps (a generalization of phase matching in
  large-etalon waveguides) possible, paving the way for efficient,
  compact, and wide-bandwdith integrated nonlinear devices.
\end{abstract}


\maketitle 

Nonlinear frequency conversion (NFC) plays a crucial role in many
photonic applications, including ultra-short pulse
shaping~\cite{DeLong94, Arbore97}, spectroscopy~\cite{Heinz82},
generating novel states of light~\cite{Kuo06,Vodopyanov06,Krischek10},
and quantum information processing~\cite{Vaziri02, Tanzilli05,
  Zaske12}.  A well-known approach for lowering the power requirements
of such nonlinear devices is that of employing optical resonators
which confine light for long times (dimensionless lifetimes $Q$) in
small volumes
$V$~\cite{JoannopoulosJo08-book,Soljacic02:bistable,Soljacic03:OL,Yanik03,Yanik04,Bravo-AbadRo10,Rivoire09,Pernice12,Bi12,Buckley14,cheng14,tang16shg}.
Although microcavity resonators designed for on-chip, infrared
applications promise some of the smallest confinement factors
available, their implementation is highly limited by the difficult
task of identifying wavelength-scale ($V \sim \lambda^3$) structures
supporting long-lived, resonant modes at widely separated wavelengths
and satisfying rigid frequency-matching and mode-overlap
constraints~\cite{Rodriguez07:OE,Bravo-AbadRo10}. Recently, we
proposed a computational framework based on large-scale
topology-optimization (TO) techniques that enables automatic discovery of
multilayer and grating structures exhibiting some of the largest SHG
figures of merit ever predicted~\cite{LinTO16}. 

In this letter, we extend our TO formulation to allow the possibility
of more sophisticated nonlinear processes and apply it to the problem
of designing rotationally symmetric and slab microresonators that
exhibit high-efficiency second harmonic generation (SHG) and
sum/difference frequency generation (SFG/DFG). In particular, we
demonstrate multi-track ring resonators and proof-of-principle
two-dimensional slab cavities supporting multiple, resonant modes
(even several octaves apart) that would be impossible to design ``by
hand''. Our designs ensure frequency matching, long radiative
lifetimes, and small (wavelength-scale) modal confinement while also
simultaneously maximizing the nonlinear modal overlap (or ``phase
matching'') necessary for efficient NFC. For instance, we discover
topology-optimized concentric ring cavities exhibiting SHG
efficiencies as high as $P_2/P_1^2=1.3\times 10^{25}
\left(\chi^{(2)}\right)^2 [\mathrm{W}^{-1}]$ even with low operational
$Q \sim 10^4$, a performance that is on a par with recently fabricated
$60\mathrm{\mu m}$-diameter, ultrahigh $Q\sim 10^6$ AlN microring
resonators~\cite{tang16shg} ($P_2/P_1^2 \sim 1.13\times 10^{24}
\left(\chi^{(2)}\right)^2 [\mathrm{W}^{-1}]$); essentially, our
topology-optimized cavities not only possess the smallest possible
modal volumes $\sim (\lambda/n)^3$, but can also operate over wider
bandwidths by virtue of their increased nonlinear modal overlap.

As reviewed in \citesasnoun{Jensen11,Liang13,LinTO16}, a typical
topology optimization problem seeks to maximize or minimize an
objective function $f$, subject to certain constraints $g$, over a set
of free variables or degrees of freedom (DOF):
\begin{align}
  \text{max}/\text{min}\, &f(\epsbar_\alpha) \\
  &g(\epsbar_\alpha) \le 0 \\
  &0 \le \epsbar_\alpha \le 1
\end{align}
where the DOFs are the normalized dielectric constants $\epsbar_\alpha
\in [0, 1]$ assigned to each pixel or voxel (indexed $\alpha$) in a
specified volume. The subscript $\alpha$ denotes appropriate spatial
discretization $\vecr \rightarrow (i,j,k)_\alpha \Delta$ with respect
to Cartesian or curvilinear coordinates. Depending on the choice of
background (bg) and structural materials, $\epsbar_\alpha$ is mapped
onto position-dependent dielectric constant via $\e_\alpha = \lp \e -
\e_\text{bg} \rp \epsbar_\alpha + \e_\text{bg}$. The binarity of the
optimized structure is enforced by penalizing the intermediate values
$\epsbar \in (0,1)$ or utilizing a variety of filter and
regularization methods~\cite{Jensen11}. Starting from a random initial
guess or completely uniform space, the technique discovers complex
structures automatically with the aid of powerful gradient-based
algorithms such as the method of moving asymptotes
(MMA)~\cite{Svanberg02}. For an electromagnetic problem, $f$ and $g$
are typically functions of the electric $\vec{E}$ or magnetic
$\vec{H}$ fields integrated over some region, which are in turn
solutions of Maxwell's equations under some incident current or
field. In what follows, we exploit direct solution of Maxwell's
equations,
\begin{align}
  \nabla \times {1 \over \mu}~\nabla \times \vec{E} -~
  \epsilon(\mathbf{r}) \omega^2 \vec{E} = i \omega \mathbf{J},
\label{eq:ME}
\end{align}
describing the steady-state $\vec{E}(\vecr;\w)$ in response to
incident currents $\vec{J}(\vecr,\w)$ at frequency $\w$.  While
solution of \eqref{ME} is straightforward and commonplace, the key to
making optimization problems tractable is to obtain a fast-converging
and computationally efficient adjoint formulation of the
problem~\cite{Jensen11}. Within the scope of TO, this requires
efficient calculations of the derivatives ${\partial f \over \partial
  \epsbar_\alpha},~{\partial g \over \partial \epsbar_\alpha}$ at
every pixel $\alpha$, which we perform by exploiting the
adjoint-variable method (AVM)~\cite{Jensen11}.


Any NFC process can be viewed as a frequency mixing scheme in which
two or more \emph{constituent} photons at a set of frequencies
$\{\w_n\}$ interact to produce an output photon at frequency
$\Omega=\sum_n c_n \w_n$, where $\{c_n\}$ can be either negative or
positive, depending on whether the corresponding photons are created
or destroyed in the process~\cite{Boyd92}. Given an appropriate
nonlinear tensor component $\chi_{ijk...}$, with
$i,j,k,...\in\{x,y,z\}$, mediating an interaction between the
polarization components $E_i(\Omega)$ and $E_{1j}$, $E_{2k}, ...$, we
begin with a collection of point dipole currents, each at the
\emph{constituent} frequency $\w_n,~n\in\{1,2,...\}$ and positioned at
the center of the computational cell $\mathbf{r}'$, such that
$\mathbf{J}_n = \hat{\mathbf{e}}_{n \nu}
\delta(\mathbf{r}-\mathbf{r}')$, where $\hat{\mathbf{e}}_{n \nu}
\in\{\hat{\mathbf{e}}_{1j},~\hat{\mathbf{e}}_{2k}, ...\}$ is a
polarization vector chosen so as to excite the desired electric-field
polarization components ($\nu$) of the corresponding mode. Given the
choice of incident currents $\mathbf{J}_n$, we solve Maxwell's
equations to obtain the corresponding \emph{constituent}
electric-field response $\vec{E}_n$, from which one can construct a
nonlinear polarization current $\mathbf{J}(\Omega) =
\bar{\epsilon}(\mathbf{r}) \prod_{n} E_{n\nu}^{|c_n| (*)}
\hat{\mathbf{e}}_i$, where $E_{n\nu}= \vE_n \cdot
\hat{\mathbf{e}}_{n\nu}$ and $\mathbf{J}(\Omega)$ can be generally
polarized ($\hat{\mathbf{e}}_i$) in a (chosen) direction that differs
from the constituent polarizations $\hat{\mathbf{e}}_{n\nu}$.  Here,
(*) denotes complex conjugation for negative $c_n$ and no conjugation
otherwise.  Finally, maximizing the radiated power, $-
\mathrm{Re}\Big[ \int \mathbf{J}(\Omega)^* \cdot \mathbf{E}(\Omega)
  ~d\mathbf{r} \Big]$, due to $\mathbf{J}(\Omega)$, one is immediately
led to the following nonlinear topology optimization (NLTO) problem:
\begin{align}
  \text{max}_{\bar{\epsilon}} ~ f(\bar{\epsilon};\w_n) &= -
  \mathrm{Re}\Big[ \int \mathbf{J}(\Omega)^* \cdot \mathbf{E}(\Omega)
    ~d\mathbf{r} \Big], \label{eq:ps1}\\ {\cal
    M}(\bar{\epsilon},\omega_n) \mathbf{E}_n &= i \omega_n
  \mathbf{J}_n,~ \mathbf{J}_n = \hat{\mathbf{e}}_{n \nu}
  \delta(\mathbf{r}-\mathbf{r}'), \notag \\ {\cal
    M}(\bar{\epsilon},\Omega) \mathbf{E}(\Omega) &= i \Omega
  \mathbf{J}(\Omega),~ \mathbf{J}(\Omega) = \bar{\epsilon} \prod_{n}
  E_{n \nu}^{|c_n| (*)} \hat{\mathbf{e}}_i, \notag \\ {\cal
    M}(\bar{\epsilon},\omega) &= \nabla \times {1 \over \mu}~\nabla
  \times -~ \epsilon(\mathbf{r}) \omega^2, \notag
  \\ \epsilon(\mathbf{r}) &= \epsilon_\text{m} + \bar{\epsilon} ~
  \left( \epsilon_\text{d} - \epsilon_\text{m} \right),
  ~\bar{\epsilon} \in [0,1]. \notag
\end{align} 
Writing down the objective function in terms of the nonlinear
polarization currents, it follows that solution of \eqref{ps1},
obtained by employing any mathematical programming technique that
makes use of gradient information, e.g. the adjoint variable
method~\cite{Jensen11}, maximizes the nonlinear coefficient (mode
overlap) associated with the aforementioned nonlinear optical
process.

\begin{figure}
\includegraphics[width=0.95\columnwidth]{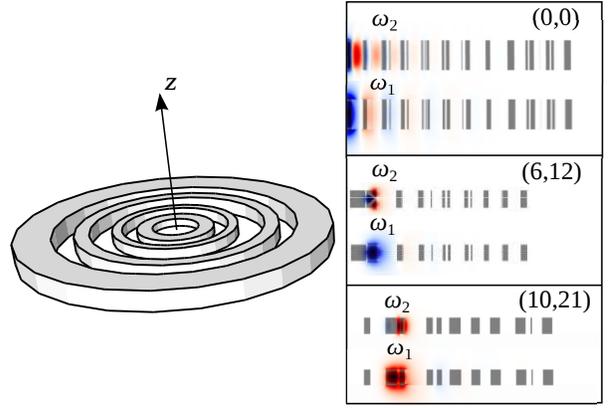}
\caption{ Schematic illustration of topology-optimized multi-track
  ring resonators. Also shown as the cross-sectional profiles of
  several ring resonators, along with those of fundamental and second
  harmonic modes corresponding to the azimuthal mode pairs (0,0),
  (6,12) and (10,21), whose increased lifetimes and modal interactions
  $\beta$ (Table~\ref{tab}) via a $\chi^{(2)}$ process lead to
  increased SHG efficiencies.\label{fig:fig1}}
\end{figure}

\begin{figure}
\includegraphics[width=0.43\textwidth]{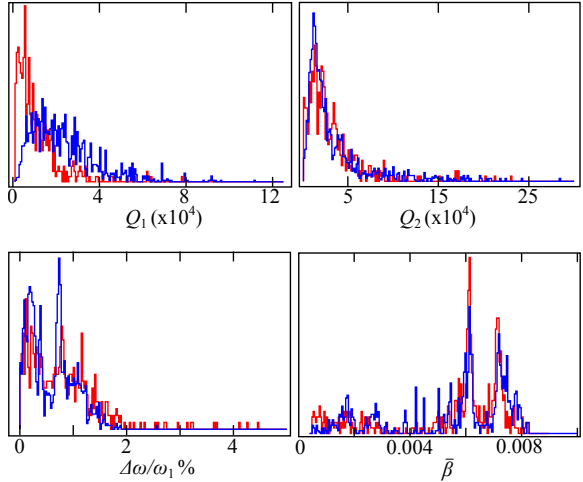}
\caption{Statistical distribution of lifetimes $Q_{1,2}$, frequency
  mismatch $\Delta \omega = |\omega_1 - \omega_2/2|$, and nonlinear
  coupling $\beta$, corresponding to the multi-track ring of
  \figref{fig1} associated with the azimuthal mode pair $(6,12)$. The
  positions of every interface is subject to random variations of
  maximum extent $\pm 36~\mathrm{nm}$ (blue line) or $\pm
  54~\mathrm{nm}$ (red line). \label{fig:fig2}}
\end{figure}

\begin{table*}
\centering
\begin{tabular}{cccccc}
\hline
$(m_1,m_2)$ & Polarization & $Q_1$             & $Q_2$             & $\bar{\beta} \left(\frac{\chi^{(2)}}{4 \sqrt(\varepsilon_0 \lambda^3)}\right)$ & Thickness $(\lambda_1)$ \\ \hline
(0, 0)      & $(E_z, E_z)$ & $10^5$            & $3\times10^4$     & 0.041          & 0.39                    \\
(4, 8)      & $(E_z, E_z)$ & $3.1\times10^4$   & $3\times 10^3$    & 0.009          & 0.30                    \\
(5, 10)     & $(E_z, E_r)$ & $8\times10^3$     & $3.7\times 10^4$  & 0.008          & 0.18                    \\
(6, 12)     & $(E_z, E_z)$ & $9.5 \times 10^4$ & $2.7 \times 10^4$ & 0.008          & 0.18                    \\
(10, 20)    & $(E_z, E_z)$ & $10^6$            & $1.2 \times 10^4$ & 0.004          & 0.22                    \\
(10, 21)    & $(E_z, E_r)$ & $1.6 \times 10^6$ & $7.4 \times 10^4$ & 0.004          & 0.24                    \\ \hline
\end{tabular}
\caption{SHG figures of merit, including azithmutal numbers $m_{1,2}$,
  field polarizations, lifetimes $Q_{1,2}$, and nonlinear coupling $\bar{\beta}$, in units of $\chi^{(2)} / 4 \sqrt(\varepsilon_0 \lambda^3)$, corresponding to the fundamental and harmonic modes of various topology-optimized multi-track ring resonators, with cross-sections (illustrated in \figref{fig1}) determined by the choice of thicknesses, given in units of $\lambda_1$.}
\label{tab}
\end{table*}

\begin{table*}
\centering
\begin{tabular}{cccccc}
\hline
$\omega_1:\omega_2:\omega_3$ & $(m_1,m_2,m_3)$ & Polarization & $(Q_1,Q_2,Q_3)$ & $\bar{\beta} \left(\frac{\chi^{(2)}}{4 \sqrt(\varepsilon_0 \lambda^3)}\right)$ & Thickness $(\lambda_1)$ \\ \hline
$1:1.2:2.2$ & $(0,0,0)$ & $(E_z,E_z,E_z)$ & $(1.8\times 10^4, 1.4 \times 10^4, 7800)$ & $0.031$ & 0.38 \\ \hline
\end{tabular}
\caption{Similar figures of merit as in Table~\ref{tab}, but for multi-track rings designed to enhance a SFG process involving light at $\w_1 = \w_3 - \w_2$, $\w_2=1.2\w_1$, and $\w_3=2.2\w_1$, with $\bar{\beta}$ described in~\citeasnoun{Rodriguez07:OE}.}
\label{tab2}
\end{table*}

\emph{Multi-track ring resonators.---} We first apply our NLTO
formulation to the design of rotationally symmetric cavities for SHG.
We consider a material platform consisting of gallium arsenide (GaAs)
thin films cladded in silica. The result of the optimizations are
described in Fig.~1 and Table~\ref{tab}, the latter of which
summarizes the most important parameters, classified according to the
choice of $m_1$ and $m_2$, which denote the azimuthal mode numbers of
fundamental and second harmonic modes, respectively. (Note that
depending on the polarization of the two modes, different
phase-matching conditions must be
imposed~\cite{Rodriguez07:OE,Bi12,cheng14}, e.g.  $m_2 = \{2m_1, 2m_1
\pm 1\}$, so in our optimizations we consider different possible
combinations.) The parameter $\bar{\beta}$ is the nonlinear coupling
strength between the interacting modes, which in the case of SHG is
given by~\cite{LinTO16}:
\begin{align}
  \bar{\beta} = { \int d\vr ~ \bar{\epsilon}(\vr) E_2^* E_1^2 \over
    \left( \int d\vr~ \epsilon_1 |\mathbf{E}_1|^2 \right) \left(
      \sqrt{\int d\vr~ \epsilon_2 |\mathbf{E}_2|^2 } \right) }
  \sqrt{\lambda_1^3 },
\label{eq:beta} 
\end{align} 
In Table.~\ref{tab2}, we also consider resonators optimized to enhance
a SFG process involving three resonant modes,
$\omega_1=\omega_3-\omega_2$, with $\omega_2=1.2\omega_1$ and
$\omega_3=2.2\omega_1$. Note that two of these modes are more than an
octave apart. The definition of the corresponding nonlinear overlap
factor, i.e. the generalization of \eqref{beta}, can be found
in~\citesasnoun{Rodriguez07:OE,burgess2009design}.

The resulting structures and figures of merit suggest the possibility
of orders of magnitude improvements. In particular, we find that the
largest overlap factors $\bar{\beta}$ are achieved in the case
$m_1=m_2=0$, corresponding to highly confined modes with peak
amplitudes near the center of the rings [Fig. 1(a)], in which case a
relatively thicker cavity $\approx 0.4 \lambda_1$ is required to
mitigate out-of-plane radiation losses. From the optimized $Q$'s and
$\bar{\beta}$ and assuming $\lambda_1 = 1.55~\mathrm{\mu m}$, we
predict a SHG efficiency of $P_2/P_1^2=1.3\times 10^{25}
\left(\chi^{(2)}\right)^2[\mathrm{W}^{-1}]$. As expected, both
radiative losses and $\bar{\beta}$ decrease with increasing $m$, as
the modes become increasingly delocalized and move away from the
center, resulting in larger mode volumes (\figref{fig1}b,c). Compared
to the state-of-the-art microring resonator demonstrated
in~\citeasnoun{tang16shg}, whose $\beta \sim 10^{-3}$, our structures
exhibit consistently larger overlaps, albeit with decreased radiative
lifetimes. The main challenge in realizing multi-track designs is
that, like photonic crystals and related structures that rely on
careful interference effects, their $Q$s tend to be more sensitive to
perturbations. In the case of centrally confined modes with
$m_1=m_2=0$, we observe the appearance of deeply subwavelength
features near the cavity center where the fields are mostly
confined. We find that these features are crucial to the integrity of
the modes since they are responsible for the delicate interference
process which cancels outgoing radiation, and therefore their absence
greatly reduces the quality factors of the modes. Overall, for
$m_1=m_2=0$, we find that for operation with $\lambda_1\sim
1.55~\mathrm{\mu m}$, a fabrication precision of several nanometers
would be necessary to ensure quality factors on the order of
$10^5$. On the other hand, the optimized designs become increasingly
robust for larger $m_1,m_2 \gg 0$ since they have fewer subwavelength
features and smaller aspect ratios. \Figref{fig2} shows distributions
of the most important figures of merit for an ensemble of
$(m_1=6,~m_2=12)$ cavities subject to random, uniformly-distributed
structural (position and thicknesses) perturbations in the range
$[-50,50]~\mathrm{nm}$. We find that while the frequency mismatch and
overlap factors are quite robust against variations, the quality
factors can decrease to $\sim 10^4$.

\begin{figure}
\includegraphics[width=0.45\textwidth]{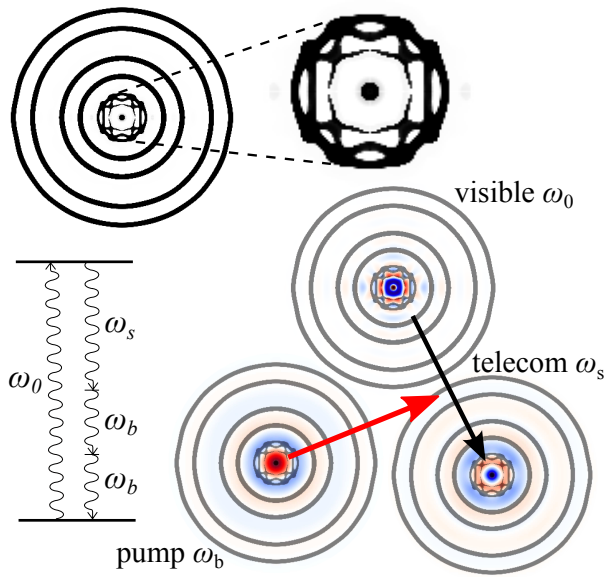}
\caption{Topology optimized 2D microcavity exhibiting tightly confined
  and widely separated modes ($\w_s,\w_b,\w_0$) that are several
  octaves apart. The modes interact strongly via a $\chi^{(3)}$ DFG
  scheme dictated by the frequency relation $\w_s=\w_0 - 2\w_b$, with
  $\w_0 = 2.35\w_s$ and $\w_b = 0.68\w_s$, illustrated by the
  accompanying two-level schematic. \label{fig:fig3}}
\end{figure}

\emph{Slab microcavities.---} We now consider a different class of
structure and NFC process, namely DFG in slab microcavities. In
particular, we consider a $\chi^{(3)}$ nonlinear process satisfying
the frequency relation $\w_s = \w_0 - 2 \w_b$, with $\w_s$, $\w_0$,
and $\w_b$ denoting the frequencies of signal, emitted, and pump
photons (see~\figref{fig3}). Such a DFG process has important
implications for single-photon frequency conversion, e.g. in
nitrogen-vacancy (NV) color centers, where a single NV photon
$\lbda_0=637~\mathrm{nm}$ is converted to a telecommunication
wavelength $\lbda_s=1550~\mathrm{nm}$ by pump light at $\lbda_b \sim
2200~\mathrm{nm}$, requiring resonances that are more than two octave
away from one another~\cite{Zin15}. In other words, the challenge is
to design a diamond cavity $(n\approx 2.4)$ that exhibits three widely
separated strongly confined modes with large nonlinear interactions
and lifetimes. \Figref{fig3} presents a proof-of-concept 2D design
that satisfies all of these requirements. Extension to 3D slabs of
finite thickness (assuming similar lateral profiles and vertical
confinement $\sim$ wavelength), one is led to the possibility of
ultra-large $\bar{\beta} \sim 0.2$, with
\begin{align}
  \bar{\beta} &= { \int d\vr ~ \bar{\epsilon}(\vr) E_0^* E_b^2 E_s
    \over \sqrt{\int d\vr~ \epsilon_0 |\mathbf{E}_0|^2} \sqrt{\int
      d\vr~ \epsilon_s |\mathbf{E}_s|^2 } \left( \int d\vr~ \epsilon_b
      |\mathbf{E}_b|^2 \right) }\lambda_1^3
\label{eq:beta}
\end{align}

Note that the lifetimes of these 2D modes are bounded only by the
finite size of our computational cell (and hence are ignored in our
discussion), whereas in realistic 3D microcavities, they will be
limited by vertical radiation losses~\cite{JoannopoulosJo08-book}.
Despite the two-dimensional aspect of this slab design, and in
contrast to the fully 3D multi-track ring resonators above, these
results provide proof of the existence of wavelength-scale photonic
structures that can greatly enhance challenging NFC processes. One
example is the NV problem described above, which is particularly
challenging if a monolithic all-diamond approach is desired, in which
case both single-photon emission and wavelength conversion are to be
seamlessly realized in the same diamond cavity~\cite{Zin15}. A viable
solution that was recently proposed is the use of four-wave mixing
Bragg scattering (FWM-BS) by way of whispering gallery
modes~\cite{Zin15,Kartik16}, which are relatively easy to phase-match
but suffer from large mode volumes. Furthermore, FWM-BS requires two
pump lasers, at least one of which has a shorter wavelength than the
converted signal photon, which could lead to spontaneous
down-conversion and undesirable noise, degrading quantum fidelity, in
contrast to the DFG scheme above, based on a long-wavelength
pump~\cite{Zin15}.

\bibliography{opt}

\end{document}